\documentclass[letterpaper,preprint]{jpsj3}
\usepackage{txfonts}
\usepackage{graphicx}
\usepackage{latexsym}
\usepackage{amsmath}
\usepackage{txfonts}
\usepackage{helvet}
\usepackage{braket}
\usepackage{amscd}
\usepackage{amssymb}
\usepackage{longtable}

\title{Local Observations of Orbital Diamagnetism and Excitation in Three-Dimensional Dirac Fermion Systems Bi$_{1-x}$Sb$_x$}

\author{Yukihiro Watanabe$^1$, Masashi Kumazaki$^1$, Hiroki Ezure$^2$, Takao Sasagawa$^2$, Robert Cava$^3$, Masayuki Itoh$^1$ and Yasuhiro Shimizu$^{1}$\thanks{yasuhiro@iar.nagoya-u.ac.jp}}
\inst{$^1$Department of Physics, Nagoya University, Nagoya, 464-8602 Japan \\
$^2$Department of Physics, Tokyo Institute of Technology, Yokohama, 226-8503 Japan \\
$^3$Department of Chemistry, Princeton University, Princeton, NJ, 08544, USA} 

\abst{Dirac fermions display a singular response against magnetic and electric fields. A distinct manifestation is large diamagnetism originating in the interband effect of Bloch bands, as observed in bismuth alloys. Through $^{209}$Bi NMR spectroscopy, we extract diamagnetic orbital susceptibility inherent to Dirac fermions in the semiconducting bismuth alloys Bi$_{1-x}$Sb$_x$ ($x = 0.08 - 0.16$). The $^{209}$Bi hyperfine coupling constant provides an estimate of the effective orbital radius. In addition to the interband diamagnetism, Knight shift includes an anomalous temperature-independent term originating in the enhanced intraband diamagnetism under strong spin-orbit coupling. The nuclear spin-lattice relaxation rate $1/T_1$ is dominated by orbital excitation and follows cubic temperature dependence in the extensive temperature range. The result demonstrates the robust diamagnetism and low-lying orbital excitation against the small gap opening, whereas $x$-dependent spin excitation appears at low temperatures. }

\begin{document}
\maketitle
Relativistic Dirac fermions exhibit large diamagnetism at room temperature, as observed in bismuth alloys Bi$_{1-x}$Sb$_x$ \cite{Wehrli,Shoenberg} and graphite \cite{Ganguli, McClure}. Similar to supercurrent of the Meissner effect, the diamagnetism of Dirac semimetals with linearly crossing bands comes from dissipationless orbital current in thermodynamic equilibrium under magnetic field. In contrast to the Landau-Peierls diamagnetism of conducting electrons, the diamagnetism in Bi$_{1-x}$Sb$_x$ is enhanced as the chemical potential $\mu$ is located close to the Dirac point \cite{Buot, Wehrli} or inside the band gap. The interband effect of Bloch bands has solved the mystery based on the exact formula of orbital susceptibility for three-dimensional (3D) Dirac fermions \cite{Fukuyama, Ogata} and relates to the giant spin Hall effect \cite{Fuseya} observed in Bi$_{1-x}$Sb$_x$ \cite{Chi}. Since the transport properties include a significant contribution of topological surface, the enhanced diamagnetism can be a complementary bulk sensitive probe of Dirac fermions. 

 Bi$_{1-x}$Sb$_x$ is semimetallic for $x = 0$ with electron and hole Fermi pockets at the $T$ and $L$ points of the Brillouin zone \cite{Liu, Li}. The Sb substitution induces the band inversion across $x \sim 0.05$ where the system becomes Weyl semimetal showing negative magnetoresistance  \cite{Schafgans,Kim,Vu, Li}. The band gap opens in the bulk for $x>0.05$, leading to a three-dimensional (3D) topological insulator involving a gapless surface state \cite{Hsieh} with extremely high mobility and quantum oscillations \cite{Jain, Qu,Taskin}. The diamagnetism becomes largest around $x \sim 0.1$ with the band gap $2\Delta \sim 10$ meV \cite{Jain}, consistent with the interband orbital susceptibility. 

In general, the orbital susceptibility $\chi_{\rm orb}$ consists of four components: the Landau-Peierls diamagnetism $\chi_{\rm LP}$ of conduction electrons, the interband orbital susceptibility $\chi_{\rm inter}$ equivalent to the Van-Vleck susceptibility, the atomic core diamagnetism $\chi_{\rm core}$, and the geometric susceptibility $\chi_{\rm geo}$ due to the Berry phase \cite{Ogata, Gao}. In massive Dirac fermion systems such as semiconducting Bi$_{1-x}$Sb$_x$ ($x > 0.05$) with small band gap, the first term $\chi_{\rm LP}$ and the spin susceptibility $\chi_{\rm spin}$ are negligible. Thus $\chi_{\rm inter}$ is expected to dominate the large diamagnetism of Bi$_{1-x}$Sb$_x$, whereas it cannot be distinguished from the other two diamagnetic factors by bulk magnetization measurements. 

Nuclear magnetic resonance (NMR) spectroscopy in principle can extract $\chi_{\rm orb}$ components independent of temperature and detect low-lying orbital excitation \cite{Maebashi, Hirosawa, Dora, Okvatovity}. Despite the long history of the material, the comprehensive NMR study is still lacked. The $\beta$-NMR measurement shows a small negative $^8$Li$^+$ Knight shift with the unknown hyperfine interaction near the surface of Bi$_{0.9}$Sb$_{0.1}$ \cite{MacFarlane}. Furthermore, orbital fluctuations can be probed by the nuclear spin-lattice relaxation rate $1/T_1$ in Dirac and Weyl semimetals \cite{Maebashi, Okvatovity}, whereas the materials so far reported involve significant spin excitation in $1/T_1$ due to the existence of Fermi surface \cite{Yasuoka, Hirata, Wang, Tian, Koumoulis,Kitagawa, Antonenko,Young,Ren,Nisson}. 

In this Letter, we demonstrate the diamagnetic hyperfine fields and the orbital excitation by $^{209}$Bi NMR experiments in Bi$_{1-x}$Sb$_x$ for $x = 0.08 - 0.16$ under magnetic field parallel and perpendicular to the $c$ axis. By comparing the Knight shift with the magnetic susceptibility, we analyze the temperature ($T$) dependent and the anomalous $T$-invariant components. The $T$ and $x$ dependences of $\chi_{\rm orb}$ and $1/T_1$ are compared with the theoretical calculation based on the 3D Dirac fermions. 

Single crystals of Bi$_{1-x}$Sb$_x$ ($x$ = 0.08, 0.10, 0.16) were prepared by the Bridgeman method. The obtained crystals display semimetallic or narrow gap semiconducting resistivity with the band gap $\Delta = 16$ meV for $x$ = 0.1, 0.08, and $2$ meV for $x$ = 0.16 (Fig. S1). The carrier concentration was estimated as $10^{16} - 10^{17}$ cm$^{-3}$ for $x$ = 0.1 and $\simeq 10^{18}$ cm$^{-3}$ for $x = 0.16$ from the Hall resistance (Fig. S2). Magnetization was measured with a superconducting quantum interference device under the magnetic field for 0.1 -- 7 T. The $^{209}$Bi NMR Knight shift $K$ and nuclear spin-lattice relaxation rate $1/T_1$ were obtained in a static magnetic field $H$ = 9.086 T. The Fourier transformed NMR spectrum was obtained from the spin-echo signal after the rf pulses $t_{\pi/2}-\tau-t_{\pi/2}-\tau$ with the pulse duration $t_{\pi/2} = 1$ $\mu$s and the interval time $\tau = 5-20$ $\mu$s. We checked the negligible rf heating effect on spin-echo signals by reducing the pulse power. The origin of $K = (\nu - \nu_0)/\nu_0$ was calibrated from the resonance frequency of the reference Bi(NO$_3$)$_3$ aqueous solution, $\nu_0$ = 62.214 MHz, where the core $s$ electron contribution was implicitly subtracted. The nuclear magnetization relaxation recovery for the central line was fitted with a multi-exponential function for the nuclear spin $^{209}I = 9/2$ (Fig. S3) \cite{wada}. 

\begin{figure}
\begin{center}
   \includegraphics[width=85mm]{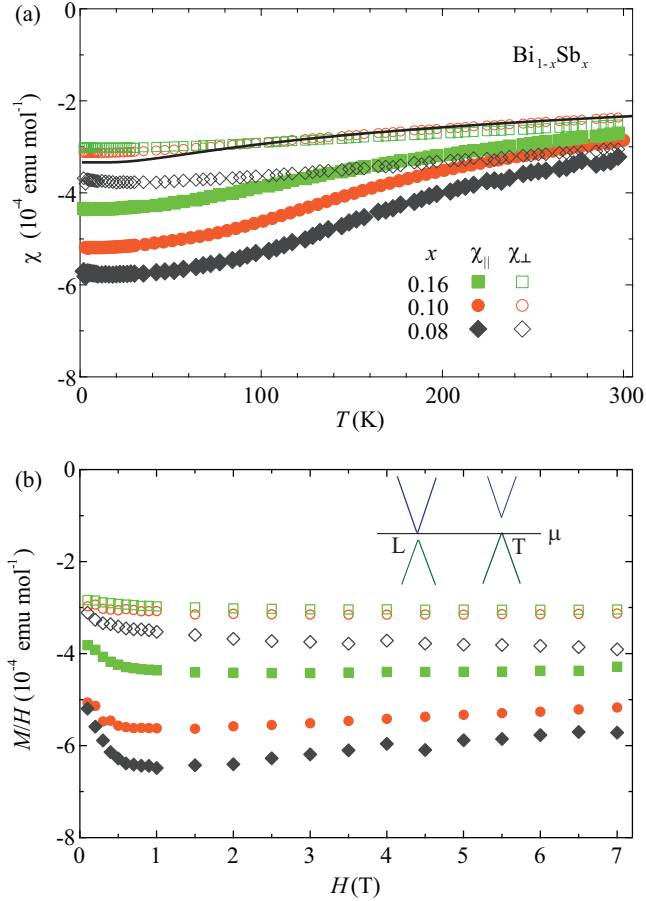}
 \caption{\label{Fig1}(a) Magnetic susceptibility $\chi$ plotted against temperature $T$ in Bi$_{1-x}$Sb$_x$ ($x = 0.08, 0.10, 0.16$) at 7 T. Magnetic field was applied parallel or normal to the $ab$ plane of the hexagonal $R{\bar 3}m$ lattice. A solid curve is the numerical calculation based on Eq.(1) for 3D Dirac fermions \cite{Maebashi}. (b) Magnetic field $H$ dependence of magnetization $M$ divided by $H$ for 0--7 T at 2.0 K. Inset: the schematic band structure around the $L$ and $T$ points.}
\end{center}
\end{figure}

Magnetic susceptibility $\chi$ was measured as a function of $T$ and $H$ normal ($\chi_\perp$) and parallel ($\chi_\parallel$) to the $ab$ plane of the hexagonal $R{\bar 3}m$ lattice in Bi$_{1-x}$Sb$_x$, as shown in Fig. 1. Here the $c$ axis is taken along the diagonal direction of the rhombohedral lattice. The calculated core diamagnetic susceptibility $-$2.3$\times$10$^{-5}$ emu mol$^{-1}$ for Bi$^{3+}$ was already subtracted. We observed anisotropic diamagnetism for three samples with the different $x$: the amplitude of $\chi_\parallel$ is greater than $\chi_\perp$ at low temperatures. The result is consistent with the theoretical calculation \cite{Fukuyama}, where the anisotropy is explained by the band structure. Namely, electron (hole) dominates the orbital motion around the $L$ ($T$) valley having nearly gapless (gapped) excitation under the magnetic field along the $a$ ($c$) axis \cite{Fukuyama, Fuseya4}. The $T$ dependence of the susceptibility is attributed to the thermal excitation comparable to a band gap energy. For $x=0.08$, $\chi_\parallel$ reaches $-5.9 \times 10^{-4}$ emu mol$^{-1}$ below 10 K. The amplitude is 2--3 times greater than the previous report \cite{Wehrli}. The diamagnetism is weakened as $x$ increases, consistent with the chemical potential $\mu$ dependence \cite{Fukuyama, Hirosawa}. 

The experimental result of $\chi_{\rm orb}$ is compared with the theoretical calculation for $\chi_{\rm inter}$ in 3D Dirac fermion systems \cite{Maebashi}. $\chi_{\rm inter}$ at finite temperatures is expressed as
\begin{equation}
\begin{split}
\chi_{\rm inter}\ =\ =-\frac{2\alpha}{3\pi}\frac{c^*}{c}\left[{\rm ln}\frac{E_\Lambda}{\Delta}-\int^\infty _\Delta d\epsilon \frac{1-f(-\epsilon)+f(\epsilon)}{\sqrt{\epsilon^2\ -\ \Delta^2}}\right], 
\end{split}    
\end{equation}
using the fine structure constant $\alpha$, $c^* \equiv \sqrt{\Delta/m^*}$ with the effective electron mass $m^*$, the band width $E_\Lambda$, and the Fermi distribution function $f(\epsilon)$ for an energy $\epsilon$. As shown in Fig. 1(a), the numerical calculation (a solid curve) using Eq.(1) qualitatively reproduces the $T$ dependence of $\chi_{\rm orb}$ for $2\Delta = 16$ meV and the energy cut off $E_\Lambda$/$\Delta$ = 400. 

The magnetization $M$ divided by $H$ weakly depends on $H$ [Fig. \ref{Fig1}(b)]. For $x$ = 0.08, $M/H$ along the $ab$ plane exhibits a minimum ($-6.6\times 10^{-4}$ emu mol$^{-1}$) around 1 T and gradually increases at high fields. Similar behavior is seen for $x$ = 0.1 and 0.16. The de Haas van Alphen oscillation due to the residual density of states \cite{Taskin} is absent in the measured field range. Although the theoretical calculation of $\chi_{\rm orb}$ under the intense field is absent, the interband effect in Eq.(1) is governed by low-lying energies ($\epsilon \sim \Delta$) and hence the transition between $n = 0$ Landau levels with increasing the magnetic field \cite{Fuseya2, Gao2, Koshino}. The population of the $n = 0$ level or $\chi_{\rm inter}$ is expected to increase with $\sqrt{H}$, which may explain the field dependent behavior of $M/H$ in a low field range. 

\begin{figure}
\begin{center}
\includegraphics[width=8.5truecm]{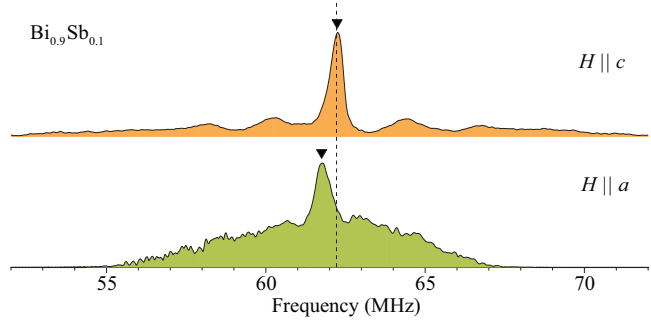}
\caption{\label{Fig2} $^{209}$Bi NMR spectrum of the Bi$_{0.9}$Sb$_{0.1}$ single crystal at 100 K under a magnetic field $H =$ 9.086 T parallel and normal to the hexagonal $c$ axis. Triangles denote the central peak position that defines  Knight shift $K$. A dotted line denotes the $K = 0$ position ($\nu_0 = 62.214$ MHz). 
}
\end{center}
\end{figure}

The $^{209}$Bi NMR spectrum (Fig. \ref{Fig2}) consisting of a sharp central line and broad satellite lines represents the first-order quadrupolar splitting in the presence of electric field gradient at the nuclear spin $^{209}I$ = 9/2. The maximum splitting along the $c$ axis gives the nuclear quadrupole frequency $\nu_{\rm Q} = 2.1$ MHz. The broadening of outer satellite lines indicates the nonuniform electric field gradient around the Sb sites. For $H \parallel c$, the central resonance frequency is located close to $\nu_0$, while it significantly shifts to a lower frequency for $H \parallel a$. The $^{209}$Bi Knight shift $K = (\nu - \nu_0)/\nu_0$ is determined from the central peak position after subtracting the second-order quadrupole contribution ($< 0.01\%$). 

 \begin{figure}
 \begin{center}
 \includegraphics[width=8.5truecm]{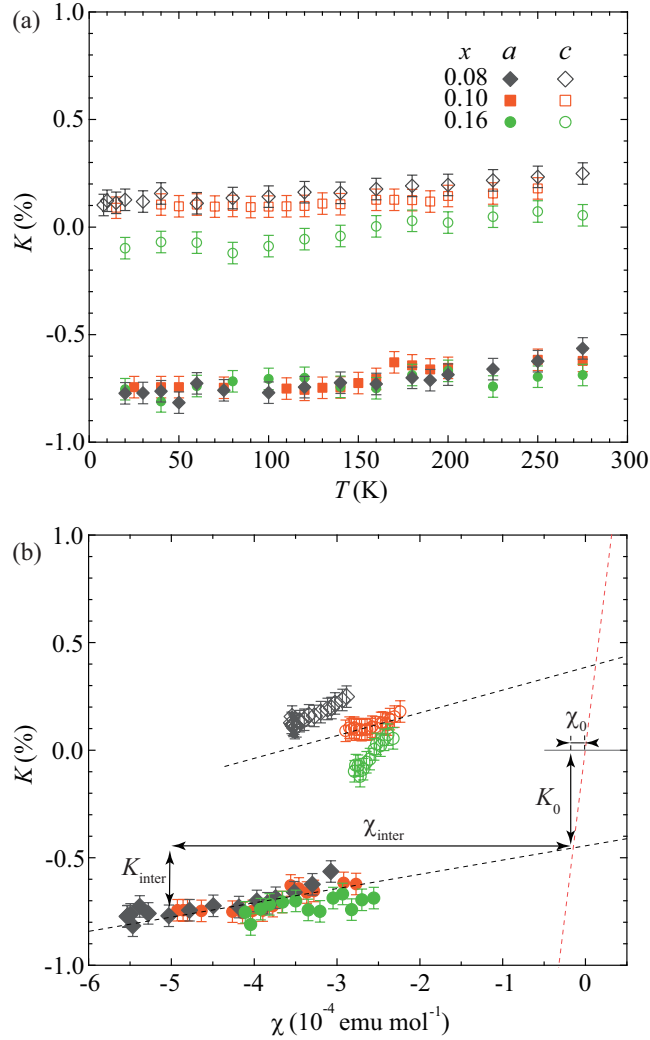}
\caption{\label{Fig3} (a) Temperature dependence of $^{209}$Bi Knight shift $K$ in Bi$_{1-x}$Sb$_x$. Magnetic field was applied along the $c$ axis (open symbols) and the $ab$ plane (filled symbols). (b) $K$ plotted against $\chi$ as an implicit function of temperature. Black and red dotted lines are the linear fitting result for $x = 0.1$ and the atomic orbital shift with the hyperfine coupling constant $A_{\rm DF}$ = 170 T/$\mu_{\rm B}$, respectively. The temperature dependent term scaling to $\chi$ is assigned to $K_{\rm inter}$. The crossing point gives the constant offsets of $\chi_0$ and $K_0$. 
}
 \end{center}
\end{figure}

The Knight shifts, $K_a$ and $K_c$, measured for $H \parallel a$ and $c$ also exhibit anisotropic behavior, as shown in Fig. 3(a). $K_a$ reaches $-0.8\%$ nearly independent of $x$, while $K_c$ remains slightly positive for $x = 0.08$ and 0.1. The diamagnetic $K_a$ is consistent with $\chi_{\parallel}$. However, the dependence of $K_a$ against $T$ is much weaker than that of $\chi_{\parallel}$. It points to a sizable $T$-independent component included in the Knight shift. The origin likely comes from the intraband orbital susceptibility, as discussed below.

Here the uniform diamagnetic shielding \cite{Maebashi} due to the bulk susceptibility $\chi_{\parallel} = - 2.4 \times 10^{-5}$ in the dimensionless unit ($x = 0.1$ at 10 K) corresponds to the Knight shift $K_{\rm dia} = 4\pi \chi = -0.031\%$ for the $a$ axis, which is an order smaller than the observed $T$-dependent component of the Knight shift. Therefore, $K$ is mostly dominated by the hyperfine interaction with the orbital angular moment of Dirac fermions. As shown in Fig. 3(b) and Table S1, the $K-\chi$ linearity gives the hyperfine coupling constant $A_{\rm DF}$ = 3.0$\pm0.5$ and 8.8$\pm 1.0$ T/$\mu_{\rm B}$ for the $a$ and $c$ axes in Bi$_{0.92}$Sb$_{0.08}$, respectively. They are much smaller than the atomic orbital hyperfine constant $A_{\rm orb} = 2N\,u_{\rm B}\braket{r^{-3}} = 170$ T/$\mu_B$ for the mean atomic radius of bismuth, $\braket{r^{-3}}$ $\sim 10^{26}$ (1/cm$^3)$ \cite{Landman}, where $N$ is the Avogadro number and $\mu_{\rm B}$ is the Bohr magneton. 
Since the mean orbital radius scales to third root of $A_{\rm DF}$, the orbital radius of Dirac fermions reaches 3--3.5 times larger than the atomic radius of bismuth for $ H \parallel a$. The $x$ dependence of $A_{\rm DF}$ along the $c$ axis may come from the difference in the trap potential of the orbital motion.

\begin{figure*}
\begin{center}
\includegraphics[width=17truecm]{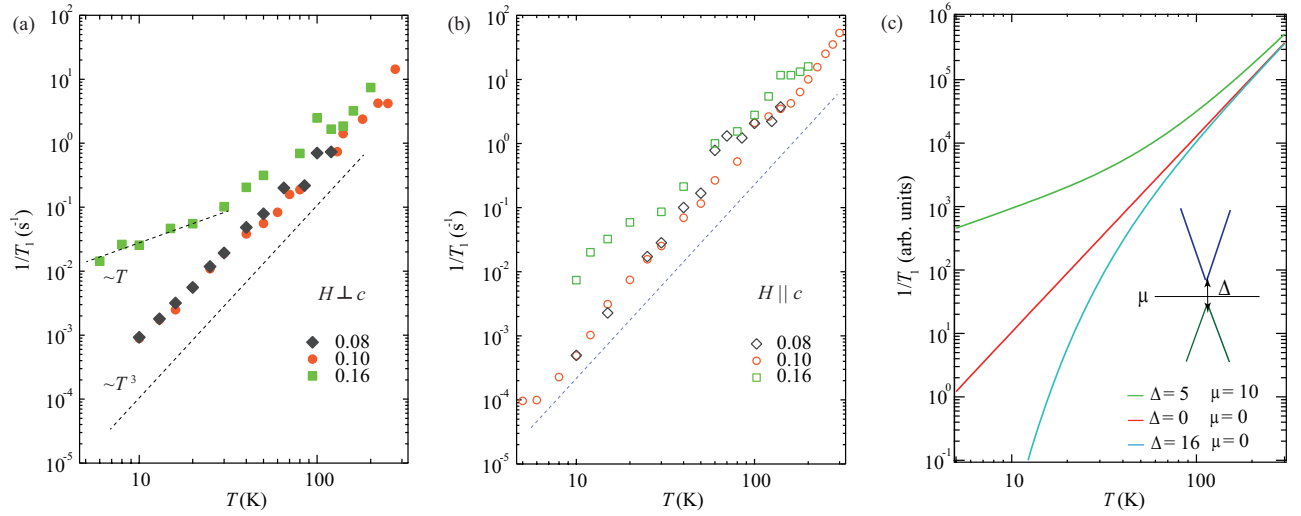}
\caption{\label{Fig4} Temperature dependence of $^{209}$Bi spin-lattice relaxation rate $1/T_1$ under the magnetic field along (a) $a$ and (b) $c$ axes of Bi$_{1-x}$Sb$_x$ ($x= 0.08$, 0.10, 0.16). (c) Calculated $1/T_1$ for 3D Dirac fermions with $\Delta = 0$ and $16$ meV for $\mu = 0$ meV, and $\Delta = 5$ meV for $\mu = 10$ meV \cite{Hirosawa}. Inset: schematic illustration of the band gap $\Delta$ and the chemical potential $\mu$ around the $L$ point. 
}
\end{center}
\end{figure*}
The low-lying magnetic excitation is investigated by the nuclear spin-lattice relaxation rate $1/T_1$ \cite{Hirosawa,Maebashi,Okvatovity}. 
As shown in Fig. \ref{Fig4}, $1/T_1$ behaves nearly isotropic and follows $T^3$ dependence for $x = 0.08$ and 0.10 in an extensive $T$ range. $1/T_1$ continues to decay more than five orders of magnitude for 5 -- 290 K. The result is consistent with the behavior expected for the intraband orbital excitation in massless 3D Dirac \cite{Hirosawa} and Weyl fermion systems \cite{Okvatovity}, which is distinct from that of the spin excitation ($1/T_1 \sim T^5$) \cite{Hirosawa}. For $x = 0.16$, $1/T_1$ deviates from the $T^3$ dependence below 100 K and becomes close to the Korriaga's law ($1/T_1 \sim T$), implying the residual density of states. 

To compare the result with the theoretical calculation for 3D Dirac fermions, we employ an expression of orbital excitation \cite{Maebashi,Tian}, 
\begin{equation}
\begin{array}{cc}\label{eq:4}
&\left(\displaystyle\frac{1}{T_1T}\right)_{\rm orb} = \displaystyle\frac{2\pi}{3}\mu_0 ^2\gamma_n ^2 e^2c^{*4}\displaystyle\int_{-\infty}^{\infty}d\epsilon \left[-\displaystyle\frac{\partial f(\epsilon,\mu)}{\partial \epsilon}\right]\displaystyle\frac{g^2(\epsilon)}{\epsilon^2}{\rm ln}\displaystyle\frac{2(\epsilon^2-\Delta^2)}{\omega_0|\epsilon|} , 
\end{array}
 \end{equation}
where $g(\epsilon)$ is the density of states. It gives $1/T_1$ $\sim T^3$ down to temperatures for $\Delta \sim 0$, as shown in Fig. 4(c), while $1/T_1$ decays exponentially in the presence of $\Delta \sim 100$ K at low temperatures. However, we observed a deviation from the $T^3$ law only below 20 K for $x = 0.08$ and 0.1 under $H \parallel c$. It suggests residual low-lying levels inside the band gap analogous to gapless excitation in the topological surface state of Bi$_{1-x}$Sb$_x$. If there is nonuniform relaxation process around the local distortion, $1/T_1$ would be spatially inhomogeneous. However, we observed only a single component down to low temperatures, indicating the uniform fluctuations distinct from the relaxation around the impurity.

The $T^3$ behavior of $1/T_1$ has been also observed in Weyl semimetals at high temperatures \cite{Yasuoka, Tian}, whereas the spin contribution becomes significant at low temperatures due to the existence of the Fermi surface. Then $1/T_1$ obeys the Korringa's relation $1/T_1T \propto K^2$ invariant against $T$. In Bi$_{1-x}$Sb$_x$, the Korringa law is grossly violated down to low temperatures owing to the predominant orbital fluctuations. The Korringa's constant $\sim 1/(T_1TK^2)$ approaches to unity only near room temperature (Fig. S4). 

Finally, we discuss the origin for the anisotropic $T$-invariant Knight shift $K_0$ deduced from the $K-\chi$ analysis in Fig. 3(b) and Table S1. For explaining the large constant term, one has to consider the other factors of intraband orbital susceptibility such as $\chi_{\rm core}$ and $\chi_{\rm geo}$, which can be enhanced under strong spin-orbit coupling ($\lambda \sim 1.5$ eV) \cite{Autschbach,Zhang,Gao}. $\chi_{\rm core}$ is isotropic for $s$ electrons but becomes anisotropic for partially filled $p$ orbitals in bismuth \cite{Ogata}. Furthermore, the contribution is enhanced by the relativistic nuclear shielding effect \cite{Autschbach}. $\chi_{\rm geo}$ due to Berry phase in Dirac fermions may also play a significant role for the anisotropic $K$ \cite{Gao, Ogata, Ominato1}. Here we consider that the Van-Vleck susceptibility identical to the interband effect gives the $T$-dependent diamagnetism for $\Delta$ comparable to the $T$ scale \cite{Ogata, Maebashi}. In addition to these orbital susceptibilities, a spin-orbital cross term $\chi_{\rm so}$ may also relate to the constant $K_0$ and $\chi_0$ in second order \cite{Ominato1,Koshino}, where the positive and negative susceptibility is expected for electron and hole bands, respectively. However, the positive (negative) $K_0$ along the $c$ ($a$) axis is inconsistent with the behavior of $\chi_{\rm so}$. Further numerical calculations are needed to explain quantitatively the anisotropy and sign of $K_0$. 

In conclusion, the orbital diamagnetism and excitation were studied through the magnetization and $^{209}$Bi NMR measurements in the single crystals of bismuth alloys Bi$_{1-x}$Sb$_x$. We observed the anisotropic and field-dependent diamagnetism in the bulk magnetic susceptibility. The temperature-dependent diamagnetic Knight shift proportional to the bulk susceptibility gives the hyperfine coupling constant of Dirac fermions with the effective orbital length scale over the unit cell. The anisotropic constant shift suggests the enhanced nuclear shielding due to strong spin-orbit coupling on bismuth. The nuclear spin-lattice relaxation rate governed by orbital current of Dirac fermions is distinct from the spin excitation following the Korringa's law. These results provide a new microscopic insight into the real-space picture of the diamagnetic orbital current in three-dimensional Dirac fermion systems.

We thank S. Inoue and T. Jinno for technical support. We are also grateful to A. Kobayashi, T. Hirosawa, and M. Ogata for fruitful discussions. This work was supported by JSPS KAKENHI (Grants No. JP19H01837, JP16H04012, and JP19H05824). 

\bibliography{bib}

\begin{thebibliography}{10}

\bibitem{Wehrli}
L.~Wehrli: Phys. Kondens. Mater. {\bfseries 8} (1968) 87.

\bibitem{Shoenberg}
D.~Shoenberg and M.~Z. Uddin: Proc. Royal Soc. London Ser. A {\bfseries 156}
  (1936) 687.

\bibitem{Ganguli}
N.~Ganguli and K.~S. Krishnan: Proc. Roy. Soc. London. A. {\bfseries 177}
  (1941) 168.

\bibitem{McClure}
J.~W. McClure: Phys. Rev. {\bfseries 104} (1956) 666.

\bibitem{Buot}
F.~A. Buot and J.~W. McClure: Phys. Rev. B {\bfseries 6} (1972) 4525.

\bibitem{Fukuyama}
H.~Fukuyama and R.~Kubo: J. Phys. Soc. Jpn. {\bfseries 28} (1970) 570.

\bibitem{Ogata}
M.~Ogata and H.~Fukuyama: Journal of the Physical Society of Japan {\bfseries
  84} (2015) 124708.

\bibitem{Fuseya}
Y.~Fuseya, M.~Ogata, and H.~Fukuyama: J. Phys. Soc. Jpn. {\bfseries 84} (2015)
  012001.

\bibitem{Chi}
Z.~Chi, Y.-C. Lau, X.~Xu, T.~Ohkubo, K.~Hono, and M.~Hayashi: Sci. Adv.
  {\bfseries 6} (2020).

\bibitem{Liu}
Y.~Liu and R.~E. Allen: Phys. Rev. B {\bfseries 52} (1995) 1566.

\bibitem{Li}
L.~Li, J.~G. Checkelsky, Y.~S. Hor, C.~Uher, A.~F. Hebard, R.~J. Cava, and
  N.~P. Ong: Science {\bfseries 321} (2008) 547.

\bibitem{Schafgans}
A.~A. Schafgans, K.~W. Post, A.~A. Taskin, Y.~Ando, X.-L. Qi, B.~C. Chapler,
  and D.~N. Basov: Phys. Rev. B {\bfseries 85} (2012) 195440.

\bibitem{Kim}
H.-J. Kim, K.-S. Kim, J.-F. Wang, M.~Sasaki, N.~Satoh, A.~Ohnishi, M.~Kitaura,
  M.~Yang, and L.~Li: Phys. Rev. Lett. {\bfseries 111} (2013) 246603.

\bibitem{Vu}
D.~M. Vu, W.~Shon, J.-S. Rhyee, M.~Sasaki, A.~Ohnishi, K.-S. Kim, and H.-J.
  Kim: Phys. Rev. B {\bfseries 100} (2019) 125162.

\bibitem{Hsieh}
D.~Hsieh, D.~Qian, L.~Wray, Y.~Xia, Y.~Hor, R.~Cava, and M.~Z. Hasan: Nature
  {\bfseries 452} (2008) 970.

\bibitem{Jain}
A.~L. Jain: Phys. Rev. {\bfseries 114} (1959) 1518.

\bibitem{Qu}
D.-X. Qu, S.~K. Roberts, and G.~F. Chapline: Phys. Rev. Lett. {\bfseries 111}
  (2013) 176801.

\bibitem{Taskin}
A.~A. Taskin and Y.~Ando: Phys. Rev. B {\bfseries 80} (2009) 085303.

\bibitem{Gao}
Y.~Gao, S.~A. Yang, and Q.~Niu: Phys. Rev. B {\bfseries 91} (2015) 214405.

\bibitem{Maebashi}
H.~Maebashi, T.~Hirosawa, M.~Ogata, and H.~Fukuyama: J. Phys. Chem. Solids
  {\bfseries 128} (2019) 138 .

\bibitem{Hirosawa}
T.~Hirosawa, H.~Maebashi, and M.~Ogata: J. Phys. Soc. Jpn. {\bfseries 86}
  (2017) 063705.

\bibitem{Dora}
B.~Dora and F.~Simon: Phys. Stat. Sol. B {\bfseries 247} (2010) 2935.

\bibitem{Okvatovity}
Z.~Okv\'atovity, F.~Simon, and B.~D\'ora: Phys. Rev. B {\bfseries 94} (2016)
  245141.

\bibitem{MacFarlane}
W.~A. MacFarlane, C.~B.~L. Tschense, T.~Buck, K.~H. Chow, D.~L. Cortie, A.~N.
  Hariwal, R.~F. Kiefl, D.~Koumoulis, C.~D.~P. Levy, I.~McKenzie, F.~H. McGee,
  G.~D. Morris, M.~R. Pearson, Q.~Song, D.~Wang, Y.~S. Hor, and R.~J. Cava:
  Phys. Rev. B {\bfseries 90} (2014) 214422.

\bibitem{Yasuoka}
H.~Yasuoka, T.~Kubo, Y.~Kishimoto, D.~Kasinathan, M.~Schmidt, B.~Yan, Y.~Zhang,
  H.~Tou, C.~Felser, A.~P. Mackenzie, and M.~Baenitz: Phys. Rev. Lett.
  {\bfseries 118} (2017) 236403.

\bibitem{Hirata}
M.~Hirata, K.~Ishikawa, G.~Matsuno, A.~Kobayashi, K.~Miyagawa, M.~Tamura,
  C.~Berthier, and K.~Kanoda: Science {\bfseries 358} (2017) 1403.

\bibitem{Wang}
C.~G. Wang, Y.~Honjo, L.~X. Zhao, G.~F. Chen, K.~Matano, R.~Zhou, and G.-q.
  Zheng: Phys. Rev. B {\bfseries 101} (2020) 241110.

\bibitem{Tian}
Y.~Tian, N.~Ghassemi, and J.~H. Ross: Phys. Rev. B {\bfseries 100} (2019)
  165149.

\bibitem{Koumoulis}
D.~Koumoulis, R.~E. Taylor, J.~McCormick, Y.~N. Ertas, L.~Pan, X.~Che, K.~L.
  Wang, and L.-S. Bouchard: J. Chem. Phys. {\bfseries 147} (2017) 084706.

\bibitem{Kitagawa}
S.~Kitagawa, K.~Ishida, M.~Oudah, J.~N. Hausmann, A.~Ikeda, S.~Yonezawa, and
  Y.~Maeno: Phys. Rev. B {\bfseries 98} (2018) 100503.

\bibitem{Antonenko}
A.~O. Antonenko, E.~V. Charnaya, D.~Y. Nefedov, D.~Y. Podorozhkin, A.~V. Uskov,
  A.~S. Bugaev, M.~K. Lee, L.~J. Chang, S.~V. Naumov, Y.~A. Perevozchikova,
  V.~V. Chistyakov, J.~C.~A. Huang, and V.~V. Marchenkov: Phys. Solid State
  {\bfseries 59} (2017) 2331.

\bibitem{Young}
B.-L. Young, Z.-Y. Lai, Z.~Xu, A.~Yang, G.~D. Gu, Z.-H. Pan, T.~Valla, G.~J.
  Shu, R.~Sankar, and F.~C. Chou: Phys. Rev. B {\bfseries 86} (2012) 075137.

\bibitem{Ren}
Z.~Ren, A.~A. Taskin, S.~Sasaki, K.~Segawa, and Y.~Ando: Phys. Rev. B
  {\bfseries 82} (2010) 241306.

\bibitem{Nisson}
D.~M. Nisson, A.~P. Dioguardi, P.~Klavins, C.~H. Lin, K.~Shirer, A.~C.
  Shockley, J.~Crocker, and N.~J. Curro: Phys. Rev. B {\bfseries 87} (2013)
  195202.

\bibitem{wada}
S.~Wada, R.~Aoki, and O.~Fujita: J. Phys. F: Met. Phys. {\bfseries 14} (1984)
  1515.

\bibitem{Fuseya4}
Y.~Fuseya, M.~Ogata, and H.~Fukuyama: J. Phys. Soc. Jpn. {\bfseries 83} (2014)
  074702.

\bibitem{Fuseya2}
Y.~Fuseya, M.~Ogata, and H.~Fukuyama: J. Phys. Soc. Jpn. {\bfseries 81} (2012)
  093704.

\bibitem{Gao2}
Y.~Gao and Q.~Niu: Proc. Nat. Acad. Sci. {\bfseries 114} (2017) 7295.

\bibitem{Koshino}
M.~Koshino and T.~Ando: Phys. Rev. B {\bfseries 81} (2010) 195431.

\bibitem{Landman}
D.~A. Landman and A.~Lurio: Phys. Rev. A {\bfseries 1} (1970) 1330.

\bibitem{Autschbach}
J.~Autschbach and T.~Ziegler: {\em Relativistic Computation of NMR Shieldings
  and Spin-Spin Coupling Constants} (John Wiley and Sons, Chichester, 2007),
  Vol.~9, pp. 303--326.

\bibitem{Zhang}
X.~Zhang, Z.~Hou, Y.~Wang, G.~Xu, C.~Shi, E.~Liu, X.~Xi, W.~Wang, G.~Wu, and
  X.-x. Zhang: Sci. Rep. {\bfseries 6} (2016) 23172.

\bibitem{Ominato1}
Y.~Ominato and K.~Nomura: Phys. Rev. B {\bfseries 97} (2018) 245207.

\end{thebibliography}

\end{document}